# Power law analysis for temperature dependence of magnetocrystalline anisotropy constants of $Nd_2Fe_{14}B$ magnets

Daisuke Miura and Akimasa Sakuma

*Department of Applied Physics, Tohoku University, Sendai 980-8579, Japan*[*]

(Dated: July 14, 2018)

We perform phenomenological analysis of the temperature dependence of magnetocrystalline anisotropy (MA) in rare-earth magnets. We define the phenomenological power laws applicable to compound magnets using the Zener theory, and we apply these laws to study the magnetocrystalline anisotropy constants (MACs) of $Nd_2Fe_{14}B$ magnets. The results indicate that the MACs closely obey the power law, and further, our analysis yields a better understanding of the temperature-dependent MA in rare-earth magnets. Furthermore, to examine the validity of the power law, we discuss the temperature dependence of the MACs in $Dy_2Fe_{14}B$ and $Y_2Fe_{14}B$ magnets as examples of cases wherein it is difficult to interpret the MA using the power law.

## I. INTRODUCTION

The complex behaviors of magnetic anisotropy (MA) in rare-earth permanent magnets have attracted considerable research attention, since such magnets are widely used as high-performance magnets that combine high magnetocrystalline anisotropy with reasonable magnetization and Curie temperature.[1–4] Typical examples of such magnets include $R_2Fe_{14}B$ magnets (where $R$ denotes a rare-earth element), which form the focus of the present study. The MA of such magnets strongly depends on temperature and is strongly influenced by the choice of $R$.[1,5–7] From previous studies, it has been found that the localized-moment picture is suitable to understand the magnetism of $R_2Fe_{14}B$ magnets. For example, Yamada et al. showed through their systematic analysis of the magnetization process for the $R_2Fe_{14}B$ series that parameters representing crystalline electric fields (CEFs) and effective exchange fields (EXFs) at $R$ sites do not depend significantly on $R$.[8] Furthermore, we have previously confirmed that an effective Hamiltonian using the parameters suitably describes the experimental temperature dependence of the magnetocrystalline anisotropy constants (MACs) of $Nd_2Fe_{14}B$ magnets.[9,10] However, there remains a gap between our microscopic picture and the macroscopic aspects of MA since previous theoretical studies on the temperature dependence of MA for $R_2Fe_{14}B$ magnets are based on complete diagonalization of the Hamiltonian[9,10] or numerical methods such as the Monte Carlo method.[4,11,12]

An effective means of bridging this gap is to develop a power-law theory[5,6,13–18] to establish the relationship between MACs and magnetization. The history of investigations on MA based on the power-law theory has been reviewed by Callen and Callen.[16] In their review article, they provided a quantum-mechanical foundation to the power-law theory and extended the foundation to more general cases such as the interaction between two moments. In deriving these power laws, the important assumptions are that a local magnetic moment picture is valid and that the magnetic structure is spatially uniform as assumed in Zener's original work[14] (in addition, a low temperature assumption is required from a mean field perspective[16,19]). However, the total magnetic moment of $R_2Fe_{14}B$ magnets is the sum of the $R$ moments and the Fe moments; thus, in the case that the directions of both moments are mutually non-collinear, the temperature dependence of the MACs will not obey the power law, and we can no longer regard the magnetic structure as a uniform one. In this regard, $Dy_2Fe_{14}B$ magnets are typical examples exhibiting high non-collinearity because of the antiferromagnetic coupling existing between the Dy and Fe moments,[1,18] and further, a wide plateau has been found in the temperature-dependence curve of their first-order MAC in the low-temperature range.[20,21] In contrast, in $Nd_2Fe_{14}B$ magnets, it is possible to assume that the Nd moments are collinear with the Fe moments.[18,22]

In this article, we aim to provide a simple and general understanding of the temperature dependence of the MA of rare-earth permanent magnets, considering $Nd_2Fe_{14}B$ magnets as an example, using the power-law theory for MACs as developed by Zener.[14] Here, we note that the original theory targets single-element magnets rather than compounds magnets; therefore, it is required to suitably correct the power law to apply it to $Nd_2Fe_{14}B$ magnets. In addition, we emphasize that the presence of a large higher–order MAC at the zero temperature considerably effects on the temperature dependence of the lower–order MACs than it; the $Nd_2Fe_{14}B$ magnet is in this case. Subsequently, we first present an explicit form of the extended power law picture within the framework of the Zener theory; next, we prove the validity of the extended power law and discuss the limitations of our approach. In addition, we briefly discuss the applicability of this law to $Y_2Fe_{14}B$ and $Dy_2Fe_{14}B$ magnets.

## II. POWER LAW FOR MACS IN TETRAGONAL SYMMETRY

$R_2Fe_{14}B$ magnets have tetragonal symmetry, and thus, their free-energy density, $\mathcal{F}(\Theta, \Phi; T)$, can be expressed as $\sum_{m=0}^{\infty} \sum_{n=0}^{\lfloor m/2 \rfloor} K_m^{(\prime)n}(T) \sin^{2m}\Theta \cos 4n\Phi$, where $K_m^{(\prime)n}(T)$ denotes the $m$th-order MAC at temperature $T$, and $\Theta$ and $\Phi$ denote the zenithal and azimuthal angles of the total magnetization direction, respectively.[1–3] This expansion is an infinite series in a purely mathematical analysis,[23] but a finite-order form is usually assumed for practical purposes. In this study, we consider the third-order form as follows:

$$\mathcal{F}(\Theta, \Phi; T) = K_1(T) \sin^2 \Theta \\ + [K_2(T) + K_2'(T) \cos 4\Phi] \sin^4 \Theta \\ + [K_3(T) + K_3'(T) \cos 4\Phi] \sin^6 \Theta, \quad (1)$$

where the angle-independent term, $K_0(T)$, is omitted. In general, $\{K_m^{(\prime)^n}(T)\}$ includes not only the $R$- and the Fe-sublattice contributions but also the cross effect between them. However, if the directions of the $R$ and Fe magnetizations are approximately collinear for arbitrary values of $\Theta$ and $\Phi$, then one can regard $\Theta$ and $\Phi$ as the angles of the $R$ or/and Fe magnetizations, and therefore, we can divide the total MA into the $R$- and Fe-sublattice contributions as

$$K_m^{(\prime)^n}(T) = K_m^{R(\prime)^n}(T) + K_m^{\mathrm{Fe}(\prime)^n}(T). \quad (2)$$

Here, by assuming that the localized $R$ moments are established by an effective exchange field induced by the Fe magnetization and that the origin of the $R$ anisotropy is local at each $R$ ion site, $K_m^{R(\prime)^n}(T)$ can be phenomenologically described by means of the Zener theory. Namely, when the free-energy density of the $R$ sublattice is expressed as $\mathcal{F}_R(\Theta, \Phi; T) \equiv \sum_{k,q} A_k^q(T) Z_k^q(\Theta, \Phi)$ in terms of a phenomenological real parameter $A_k^q(T)$ and the tesseral harmonics $Z_k^q(\Theta, \Phi)$, the Zener theory[14,15] yields the Akulov–Zener power law, $A_k^q(T) = A_k^q(0) \mu_R(T)^{k(k+1)/2}$, where $\mu_R(T)$ denotes the normalized $R$ magnetization as $\mu_R(0) = 1$. On the other hand, $\{A_k^q(T)\}$ is related to $\{K_m^{R(\prime)^n}(T)\}$ by another form corresponding to Eq. (1):

$$\mathcal{F}_R(\Theta, \Phi; T) = \sum_{m=1}^{3} \sum_{n=0}^{\lfloor m/2 \rfloor} K_m^{R(\prime)^n}(T) \sin^{2m} \Theta \cos 4n\Phi, \quad (3)$$

and thus, we obtain the below equations of interest:

$$K_1^R(T) = K_1^R(0) \mu_R(T)^3 \\ + \frac{8}{7} K_2^R(0) \left[ \mu_R(T)^3 - \mu_R(T)^{10} \right] \\ + \frac{8}{7} K_3^R(0) \left( \mu_R(T)^3 - \frac{18}{11} \mu_R(T)^{10} + \frac{7}{11} \mu_R(T)^{21} \right), \quad (4a)$$

$$K_2^R(T) = K_2^R(0) \mu_R(T)^{10} \\ + \frac{18}{11} K_3^R(0) \left[ \mu_R(T)^{10} - \mu_R(T)^{21} \right], \quad (4b)$$

$$K_2^{R\prime}(T) = K_2^{R\prime}(0) \mu_R(T)^{10} \\ + \frac{10}{11} K_3^{R\prime}(0) \left[ \mu_R(T)^{10} - \mu_R(T)^{21} \right], \quad (4c)$$

$$K_3^R(T) = K_3^R(0) \mu_R(T)^{21}, \quad (4d)$$

$$K_3^{R\prime}(T) = K_3^{R\prime}(0) \mu_R(T)^{21}, \quad (4e)$$

in which the first term represents the Akulov–Zener–Callen–Callen power law (or today simply called the Callen–Callen law),[13,14,16] and the second and/or the third terms represent the corrections to the Callen–Callen law due to the presence of the higher–order MAC (s) at the zero temperature.

Here, we should notice a connection between Zener's phenomenological theory and a microscopic theory. The Zener theory has been discussed well by using a mean field theory (MFT),[16,19,24–26] and the present phenomenological expressions (4) can be supported within a linear theory for CEFs. Yamada et al. explicitly represented $K_m^{R(\prime)^n}(0)$ in terms of the CEF parameters by using the perturbative theory with respect to the CEF.[8] Recently, on the basis of the expression, an MA has been discussed from first principles[4,27–35] although in general, the quantitative validity of the perturbative treatment for the MA of rare-earth magnets is not trivial.[8] In finite temperatures, Kuz'min represented $K_m^{R(\prime)^n}(T)$ in terms of the CEF and the EXF within the same framework, where the temperature dependence of the MACs is related to Kuz'min's generalized Brillouin functions (GBFs),[24,26,36,37] and furthermore, Magnani et al. investigated the effect of a finite spin–orbit interaction on $K_m^{R(\prime)^n}(T)$.[38] As pointed out by Keffer,[19] the temperature dependence of MA predicted by the MFT tends to the Akulov–Zener power law in a low temperature range, and in fact, it has been demonstrated that the GBFs and the power law are in quantitative agreement in the low temperature range.[18,26] Although the MFT does not support the power law quantitatively in a high temperature range, it is worth enough to employ the power law in the whole temperature range if limiting the discussion to a qualitative level.

To understand the temperature dependence of the MA represented by the present power law (4), let us consider a simple case wherein $K_2^{R\prime}(0) = K_3^R(0) = K_3^{R\prime}(0) = 0$:

$$K_1^R(T) = K_1^R(0) \mu_R(T)^3 + \frac{8}{7} K_2^R(0) \left[ \mu_R(T)^3 - \mu_R(T)^{10} \right], \quad (5a)$$

$$K_2^R(T) = K_2^R(0) \mu_R(T)^{10}. \quad (5b)$$

A schematic of $\mu_R(T)$ and the resultant $\mu_R(T)^3$, $\mu_R(T)^{10}$, and $\frac{8}{7} \left[ \mu_R(T)^3 - \mu_R(T)^{10} \right]$ are shown in Fig. 1. In the case that $\left| K_2^R(0)/K_1^R(0) \right| \ll 1$, the correction resulting from the presence of $K_2^R(0)$ in Eq. (5a) is small, and we confirm the applicability of the Callen–Callen law $K_1^R(T) \simeq K_1^R(0) \mu_R(T)^3$. An interesting feature is observed in the case where $K_2^R(0)$ is comparable to $K_1^R(0)$. In this case, the correction to $K_1^R(T)$ from $K_2^R(0)$ immediately occurs on increasing the temperature from absolute zero, as indicated by the dashed line in Fig. 1;

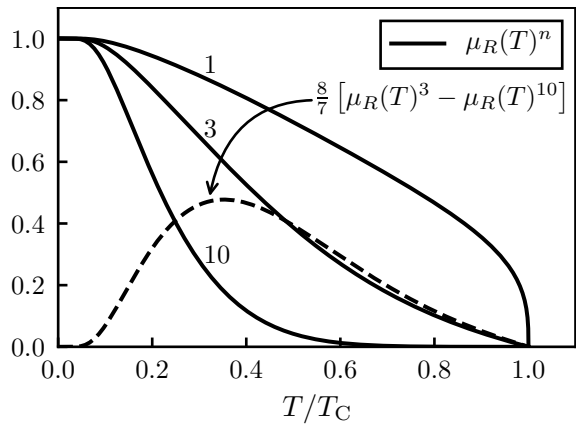

FIG. 1. Schematic of the normalized $R$ magnetization as a function of $T$, and the resultant of each component of the power law. The solid lines indicate $\mu_R(T)^n$ where $n$ is represented by the number on each solid line. The dashed line indicates the correction component of the power law. Further, $T_C$ represents the Curie temperature.

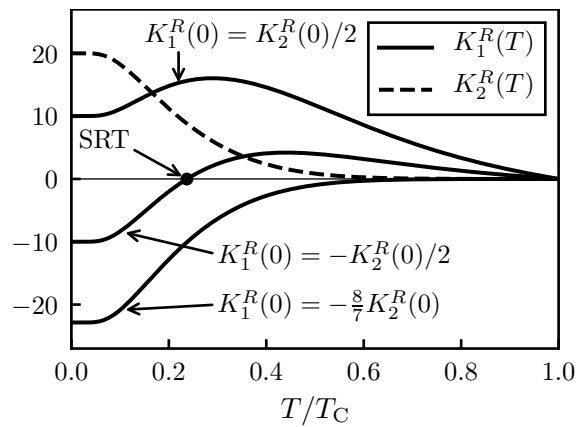

FIG. 2. Possible forms of $K_1^R(T)$ for a fixed $K_2^R(T)$ with $K_2(0) > 0$.

thus, the Callen–Callen law for $K_1^R(T)$ is no longer valid in the low-temperature range. In the high-temperature range satisfying $\mu_R(T)^3 \gg \mu_R(T)^{10}$, we can determine the power law as $K_1^R(T) \simeq \left[ K_1^R(0) + \frac{8}{7} K_2^R(0) \right] \mu_R(T)^3$ again; however, the proportional factor differs from that of the Callen–Callen law. Figure 2 illustrates the possible forms of $K_1^R(T)$ for a fixed $K_2^R(T)$ in Eqs. (5), wherein we fixed $K_2^R(T)$ as $K_2^R(0) = 20$ MJ/m$^3$ (dashed line) and determined $K_1^R(T)$, indicated by solid lines, for cases where $K_1^R(0) = K_2^R(0)/2$, $K_1^R(0) = -K_2^R(0)/2$, and $K_1^R(0) = -8K_2^R(0)/7$. From the figure, we can observe a uniaxial magnetic anisotropy over the whole temperature range in the case of $K_1^R(0) > 0$ and an in-plane magnetic anisotropy in the case of $K_1^R(0) < -8K_2^R(0)/7$. However, in the case where $K_1^R(0)$ takes on an intermediate value with $-8K_2^R(0)/7 < K_1^R(0) < 0$, we can observe that the MA changes from in-plane to uniaxial at a particular temperature; that is, spin reorientation transition (SRT) occurs at the temperature indicated by the solid line corresponding to $K_1^R(0) = -K_2^R(0)/2$ in Fig. 2. Therefore, we can immediately predict from the values of $K_1^R(0)$ and $K_2^R(0)$ as to whether the SRT occurs, although the actual SRT temperature cannot be determined because it depends on $\mu_R(T)$. As the form of $\mu_R(T)$ is approximately the same for rare-earth permanent magnets,[20] the above analysis is valid for magnets in which non-collinearity is negligible.

The application of the Zener theory to the Fe anisotropy (i.e., $K_m^{\text{Fe}(\prime)^n}(T)$ in Eq. (2)) requires caution, because it is not trivial that a local-moment picture for the Fe sublattice is appropriate. This issue is referred in a discussion on the anisotropy of Y$_2$Fe$_{14}$B magnets in the next section.

## III. TEMPERATURE DEPENDENCE OF MACS IN $R_2$Fe$_{14}$B MAGNETS

Let us apply the power law (4) to an experiment to study the temperature dependence of the MACs of Nd$_2$Fe$_{14}$B magnets. One way to estimate $\mu_R(T)$ of an $R_2$Fe$_{14}$B magnet experimentally is to subtract the magnetization of Y$_2$Fe$_{14}$B magnets from that of the $R_2$Fe$_{14}$B magnet. In this regard, Hirosawa et al. have shown that $\mu_R(T)$ estimated by this method is reproduced suitably within the mean field theory.[20] Although one can directly use the experimental $\mu_R(T)$, here, we use the theoretical $\mu_R(T)$ based on the mean field theory as follows. Assuming an effective exchange interaction between the total angular momentum of the $R$ ions and the Fe spins to plot the $\mu_R(T)$ curve, we have the conventional result:[20,26]

$$\mu_R(T) := B_J \left( \frac{2|g_{LSJ} - 1| J H_R \mu_{\text{Fe}}(T)}{k_B T} \right), \quad (6)$$

$$B_J(x) := \frac{2J+1}{2J} \coth\left( \frac{2J+1}{2J} x \right) - \frac{1}{2J} \coth\left( \frac{x}{2J} \right), \quad (7)$$

$$g_{LSJ} := 1 + \frac{J(J+1) + S(S+1) - L(L+1)}{2J(J+1)}, \quad (8)$$

where $J$ denotes the quantum number with respect to the total angular momentum of the $R$ ion, $H_R$ the effective exchange field, $k_B$ the Boltzmann constant, and $T_C$ the Curie temperature of the $R_2$Fe$_{14}$B magnet. The temperature dependent Fe magnetization is phenomenologically expressed by the Kuz'min formula:[39,40]

$$\mu_{\text{Fe}}(T) = \left[ 1 - 0.5 \left( \frac{T}{T_C} \right)^{3/2} - 0.5 \left( \frac{T}{T_C} \right)^{5/2} \right]^{1/3}, \quad (9)$$

where, the shape parameters have been determined by fitting with the temperature-dependent magnetization of the Y$_2$Fe$_{14}$B magnet reported in Ref[20,21,41]. The other parameters for Nd$_2$Fe$_{14}$B magnets are given as $L = 6$, $S = 3/2$, $J = 9/2$, $H_{\text{Nd}} = 350$ K,[8] and $T_C = 586$

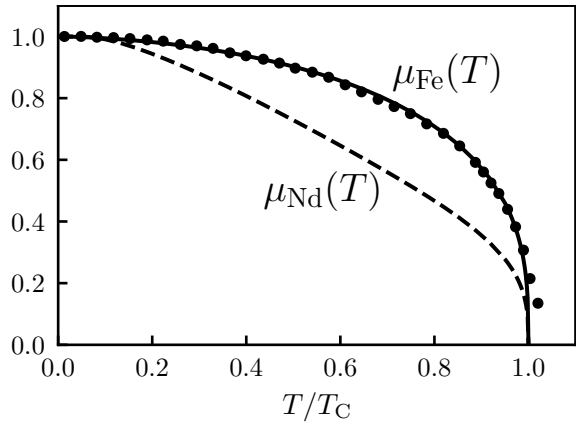

FIG. 3. Temperature dependences of the normalized magnetizations of Fe and Nd sublattices. The closed circles denote the experimental results for the $Y_2Fe_{14}B$ magnet ($T_C = 571$ K),[20,21,41] the solid line represents $\mu_{Fe}(T)$ calculated by the Kuz'min formula, and the dashed line represents $\mu_{Nd}(T)$ calculated within the framework of the molecular field theory where the effective exchange field is given by $350\mu_{Fe}(T)$ [K] and $T_C = 586$ K.

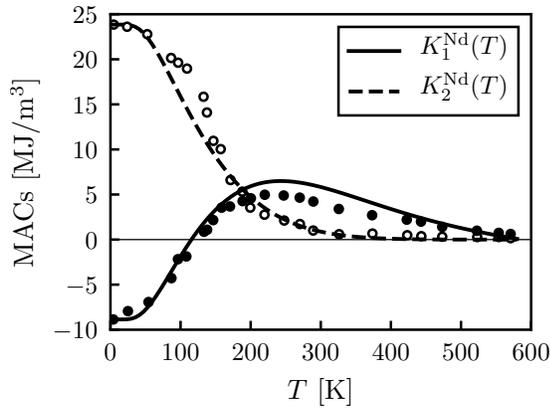

FIG. 4. Comparison of the calculated results with those of the experiment conducted by Durst et al.,[42] for the MACs of the $Nd_2Fe_{14}B$ magnet. The closed and open circles, respectively, are the first-order and the second-order MACs in the experimental results, and the lines represent results of this study.

K.[20,21] The experimental results, $\mu_{Fe}(T)$, and the resultant $\mu_{Nd}(T)$ are shown in Fig. 3. Using the above parameters, we demonstrate comparisons of our extended power law with two experiments. Figure 4 compares the calculated result with the results of the experiment conducted by Durst et al.,[42] who determined the first-order and the second-order MACs for $Nd_2Fe_{14}B$ magnets. In our study, the solid lines in the figure were obtained using Eqs. (5) and (6) where $K_1^{Nd}(0) = -8.86$ MJ/m$^3$ and $K_2^{Nd}(0) = 23.85$ MJ/m$^3$; we note that the experimental results obey the present power law qualitatively. Another example corroborating the validity of our approach is shown in Fig. 5. Cadogan et al. reported experi-

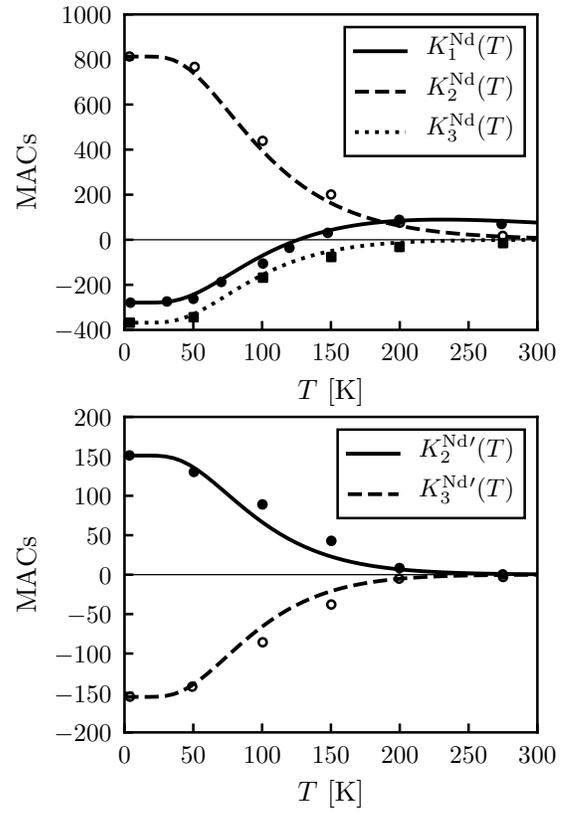

FIG. 5. Comparison of the calculated results with those of the experiment conducted by Cadogan et al.,[22] for the MACs of the $Nd_2Fe_{14}B$ magnet. In the upper figure, the closed circles, open circles, and closed rectangles represent the experimental $K_1(T)$, $K_2(T)$, and $K_3(T)$ values, respectively. In the lower figure, the closed and open circles represent the experimental $K_2'(T)$ and $K_3'(T)$ values, respectively. The lines represent the results of this study.

mental results up to third-order MACs for a $Nd_2Fe_{14}B$ magnet.[22] With the use of these results, the solid lines in the figure were plotted utilizing Eqs. (4) and (6) where $K_1^{Nd}(0) = -279$, $K_2^{Nd}(0) = 813$, $K_2^{Nd\prime}(0) = 151$, $K_3^{Nd}(0) = -368$, and $K_3^{Nd\prime}(0) = -155$ (the unit is unknown). Further, in this case, it can be observed that the present power law correlates well with the experimental results. From the results of both experiments, we can conclude that the proposed predictions and experimental results are in good agreement. Here, we emphasize that the common parameters in the two comparisons are used in $\mu_{Nd}(T)$, and particularly that the value of $H_{Nd} = 350$ K is reasonable because the strength of the EXF strongly influences the plateau of the MACs appearing in the low-temperature range.[9] In addition, we must consider the non-collinearity between the $R$ and Fe moments. Although the MACs of $R =$Nd obey the present power law well, this is not the case for magnets with a high non-collinearity. For example, $Dy_2Fe_{14}B$ magnets show a high non-collinearity between Dy and Fe moments.[18] In fact, the wide plateau observed in $K_1(T)$ up to $T \sim 0.5T_C$ from

$T = 0$[20,21] cannot be explained using the present power law for $H_{\text{Dy}} = 145$ K, as estimated by Yamada et al.[8] This violation is attributed to the high non-collinearity, not meaning that 145 K is a very small value. Therefore, a special treatment beyond the present one is necessary in such a case. A detailed analysis on $Dy_2Fe_{14}B$ magnets will be presented in a separate paper.

Finally, let us consider an interesting example in the context of the above discussion, wherein we have ignored the contribution to the MA from the Fe sublattice, $K_m^{\text{Fe}(\prime)^n}(T)$, which appeared in Eq. (2). This contribution is, often assuming that the Fe sublattice has an intrinsic MA, estimated from the MA measured in $M_2Fe_{14}B$ magnets with no $M$ moment ($M =$ Y, La, Ce, and so on).[1] Here, let us consider the $Y_2Fe_{14}B$ magnet referred to in the above discussion. The history of phenomenological investigations on the temperature-dependence of the MA of $Y_2Fe_{14}B$ magnets was reviewed by Cadogan et al.[43] In a similar manner, we attempted to apply the power-law-like expression to the temperature-dependent MACs of $Y_2Fe_{14}B$ magnets. As an analogy to Eq. (4a), we assume the following expression (previously assumed by Carr,[15] the same formulae were used in the study on the MA of hexagonal Co), as the Fe contribution to $K_1(T)$ in $Y_2Fe_{14}B$ magnets,

$$\begin{aligned} K_1^{\text{Fe}}(T) &= K_1^{\text{Fe}}(0)\mu_{\text{Fe}}(T)^3 \\ &+ \frac{8}{7}K_2^{\text{Fe}}(0)\left[\mu_{\text{Fe}}(T)^3 - \mu_{\text{Fe}}(T)^{10}\right] \\ &+ \frac{8}{7}K_3^{\text{Fe}}(0)\left[\mu_{\text{Fe}}(T)^3 - \frac{18}{11}\mu_{\text{Fe}}(T)^{10} + \frac{7}{11}\mu_{\text{Fe}}(T)^{21}\right], \end{aligned} \quad (10)$$

where we use the same parameters as the corresponding ones in Fig. 3 for $\mu_{\text{Fe}}(T)$; thus, the remaining adjustable parameters are $K_1^{\text{Fe}}(0)$, $K_2^{\text{Fe}}(0)$, and $K_3^{\text{Fe}}(0)$. Fitting the form (10) to the experimental data,[20,21] we obtain $K_1^{\text{Fe}}(0) = 0.77$ MJ/m$^3$, $K_2^{\text{Fe}}(0) = 1.21$ MJ/m$^3$, and $K_3^{\text{Fe}}(0) = 0.11$ MJ/m$^3$. The comparison of the expression (10) with the experimental data is shown in Fig. 6, wherein we note that a good fit is obtained. Here, we should emphasize that assuming the presence of finite $K_2^{\text{Fe}}(0)$ and $K_3^{\text{Fe}}(0)$ leads us to a considerably large $K_2^{\text{Fe}}(T)$ value (and, a small but not negligible $K_3^{\text{Fe}}(T)$ value) consistent with the treatment analogy using Eqs. (4b) and (4d). However, the presence of such a higher-order MAC has not thus far been reported, as mentioned in an earlier review.[43] We also plot these ghost MACs in the form of dashed and dotted lines in Fig. 6. This situation is the same as that encountered by Zener in the study on the MA of nickel,[14] which suggests that the assumptions of the Zener theory are invalid for $Y_2Fe_{14}B$ magnets. Therefore, the expression (10) along with Eq. (9) should be used only for obtaining the continuous values of $K_1^{\text{Fe}}(T)$, and not for explaining the mechanisms of the MA in the Fe sublattice; this remains an unresolved problem in itinerant electron systems such as that in the case of the temperature-dependence of the MA of

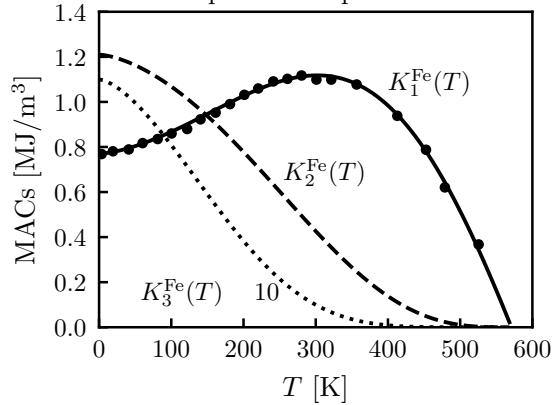

FIG. 6. Comparison of the experimental results (closed circles) with the phenomenological expression based on the Zener theory (lines). The solid lines are obtained by fitting the linear combination of $\mu_{\text{Fe}}(T)^3$, $\mu_{\text{Fe}}(T)^{10}$, and $\mu_{\text{Fe}}(T)^{21}$ to the experimental first-order MAC of the $Y_2Fe_{14}B$ magnet.[20,21] The dashed and dotted lines, respectively, represent the second-order and the third-order MACs that should be present for consistency of the Zener theory.

MnBi.[44,45]

## IV. SUMMARY

In summary, we have shown explicit forms of the power law applicable to rare-earth magnets within the framework of the Zener theory, and we have demonstrated that the power law suitably describes the experimental MACs of $Nd_2Fe_{14}B$ magnets. Thus, we obtained a simple and general understanding of the complex temperature dependence of the MA often observed in rare-earth magnets. Furthermore, we considered $Dy_2Fe_{14}B$ and $Y_2Fe_{14}B$ magnets as examples regarding the interpretation of MA when the power law failed. We believe that our findings can further contribute to studies on the magnetocrystalline anisotropy of rare-earth magnets.

## ACKNOWLEDGMENTS

This work was supported by JSPS KAKENHI Grant No. 16K06702 and 17K14800 in Japan.

---

* dmiura@solid.apph.tohoku.ac.jp
[1] J. F. Herbst, Rev. Mod. Phys. **63**, 819 (1991).
[2] R. Skomski and J. M. D. Coey, *Permanent Magnetism* (CRC Press, 1999).

3. J. M. D. Coey, "Magnetism and Magnetic Materials," (2010).
4. T. Miyake and H. Akai, J. Phys. Soc. Japan **87**, 041009 (2018), arXiv:1801.03455.
5. R. Skomski, O. N. Mryasov, J. Zhou, and D. J. Sellmyer, J. Appl. Phys. **99**, 6 (2006).
6. R. Skomski and D. J. Sellmyer, J. Rare Earths **27**, 675 (2009).
7. R. Skomski, P. Kumar, G. C. Hadjipanayis, and D. J. Sellmyer, IEEE Trans. Magn. **49**, 3229 (2013).
8. M. Yamada, H. Kato, H. Yamamoto, and Y. Nakagawa, Phys. Rev. B **38**, 620 (1988).
9. R. Sasaki, D. Miura, and A. Sakuma, Appl. Phys. Express **8**, 043004 (2015), arXiv:1501.01782.
10. D. Miura, R. Sasaki, and A. Sakuma, Appl. Phys. Express **8**, 113003 (2015), arXiv:1505.05686v1.
11. M. Matsumoto, H. Akai, Y. Harashima, S. Doi, and T. Miyake, J. Appl. Phys. **119**, 213901 (2016).
12. Y. Toga, M. Matsumoto, S. Miyashita, H. Akai, S. Doi, T. Miyake, and A. Sakuma, Phys. Rev. B **94**, 174433 (2016).
13. N. Akulov, Z. Phys. **100**, 197 (1936).
14. C. Zener, Phys. Rev. **96**, 1335 (1954).
15. W. J. Carr, Phys. Rev. **109**, 1971 (1958).
16. H. B. Callen and E. Callen, J. Phys. Chem. Solids **27**, 1271 (1966).
17. R. Skomski, J. Appl. Phys. **83**, 6724 (1998).
18. M. Ito, M. Yano, N. M. Dempsey, and D. Givord, J. Magn. Magn. Mater. **400**, 379 (2016).
19. F. Keffer, Phys. Rev. **100**, 1692 (1955).
20. S. Hirosawa, Y. Matsuura, H. Yamamoto, S. Fujimura, M. Sagawa, and H. Yamauchi, J. Appl. Phys. **59**, 873 (1986).
21. M. Sagawa, S. Hirosawa, H. Yamamoto, S. Fujimura, and Y. Matsuura, Jpn. J. Appl. Phys. **26**, 785 (1987).
22. J. M. Cadogan, J. P. Gavigan, D. Givord, and H. S. Li, J. Phys. F Met. Phys. **18**, 779 (1988).
23. M. D. Kuz'min, Phys. Rev. B **51**, 8904 (1995).
24. Y. Millev and M. Fähnle, Phys. Rev. B **51**, 2937 (1995).
25. A. A. Kazakov and R. I. Andreeva, Sov. Phys. Solid State **12**, 192 (1970).
26. M. D. Kuz'min, Phys. Rev. B **46**, 8219 (1992).
27. K. Hummler and M. Fähnle, Phys. Rev. B **53**, 3290 (1996).
28. H. Moriya, H. Tsuchiura, and A. Sakuma, J. Appl. Phys. **105**, 07A740 (2009).
29. Y. Toga, H. Moriya, H. Tsuchiura, and A. Sakuma, J. Phys. Conf. Ser. **266**, 012046 (2011).
30. S. Tanaka, H. Moriya, H. Tsuchiura, A. Sakuma, M. Diviš, and P. Novák, J. Appl. Phys. **109**, 07A702 (2011).
31. T. Yoshioka, H. Tsuchiura, and P. Novák, Mater. Res. Innov. **19**, S3 (2015).
32. Y. Tatetsu, S. Tsuneyuki, and Y. Gohda, Phys. Rev. Appl. **6**, 064029 (2016).
33. Y. Tatetsu, Y. Harashima, T. Miyake, and Y. Gohda, (2018), arXiv:1802.05817.
34. T. Yoshioka and H. Tsuchiura, Appl. Phys. Lett. **112**, 162405 (2018).
35. H. Tsuchiura, T. Yoshioka, and P. Novák, Scr. Mater. **X**, XXX (2018).
36. M. D. Kuz'min, J. Magn. Magn. Mater. **154**, 333 (1996).
37. M. D. Kuz'min and A. M. Tishin, in *Handb. Magn. Mater.*, Vol. 17, edited by K. H. J. Buschow (Elsevier, 2007) pp. 149–233.
38. N. Magnani, S. Carretta, E. Liviotti, and G. Amoretti, Phys. Rev. B **67**, 144411 (2003).
39. M. D. Kuz'min, Phys. Rev. Lett. **94**, 107204 (2005).
40. M. D. Kuz'min, D. Givord, and V. Skumryev, J. Appl. Phys. **107**, 113924 (2010).
41. S. Hirosawa, Y. Matsuura, H. Yamamoto, S. Fujimura, M. Sagawa, and H. Yamauchi, Jpn. J. Appl. Phys. **24**, L803 (1985).
42. K. D. Durst and H. Kronmuller, J. Magn. Magn. Mater. **59**, 86 (1986).
43. J. M. Cadogan and H.-S. Li, J. Magn. Magn. Mater. **110**, L15 (1992).
44. B. Roberts, Phys. Rev. **104**, 607 (1956).
45. T. Chen and W. E. Stutius, IEEE Trans. Magn. **10**, 581 (1974).